\pdfoutput=1
\documentclass[aps,prl,showpacs,twocolumn,superscriptaddress,showpacs,groupedaddress,nofootinbib]{revtex4-1}  

\usepackage{color}
\usepackage{graphicx} 
\usepackage{dcolumn} 
\usepackage{bm}       
\usepackage{amssymb}   
\usepackage{amsmath,amssymb,amsthm,mathrsfs,amsfonts,dsfont} 
\usepackage[colorlinks=true, citecolor=darkblue, linkcolor=darkblue,  urlcolor=darkblue]{hyperref}
\usepackage[dvipsnames]{xcolor}

\definecolor{darkblue}{rgb}{0.0, 0.0, 0.55}
\def\be{\begin{eqnarray}}
\def\ee{\end{eqnarray}}
\def \bea {\begin{eqnarray}}
\def \eea {\end{eqnarray}}

\def \nn {\nonumber}

\def \zb {\bar z}

\begin{document}

\title{Holographic Derivation of BPZ-Type Null-State Equations \\
\vspace{0.1cm}
in Higher Dimensional CFTs}
\author{Kuo-Wei Huang \vspace{0.1cm}}
\affiliation{
Mathematical Sciences, University of Southampton, \\
Highfield, Southampton SO17 1BJ, UK
}

\fontsize{10pt}{12pt}\selectfont

\begin{abstract}

A set of linear differential equations was recently put forward as higher-dimensional generalizations of the BPZ null-state equations in two-dimensional CFTs at large central charge. 
In this work, we derive these higher-dimensional equations from gravity, based on the AdS/CFT correspondence. 
A near-boundary expansion is employed to analyze a light scalar field equation in a black hole background. 
There is a decoupling mechanism in the bulk perturbative series at certain conformal dimensions, resulting in isolated lower-order equations. 
We find that the results agree with the previously proposed four-dimensional CFT equations, which capture the resummed contributions from minimal-twist multi-stress tensor operators. The holographic calculation also allows one to obtain additional CFT differential equations that extend beyond the near-lightcone regime. 

\end{abstract}

\maketitle

\noindent {\bf Introduction:}  
Developing analytic approaches for the study of non-perturbative physical observables, particularly in the dimensions relevant to our world, remains a major challenge in quantum field theory. The emergence of conformal invariance in critical phenomena gives rise to a special class of quantum theories -- conformal field theories (CFTs). With more rigorous mathematical structures, these theories are central to many areas of modern physics, ranging from many-body phase transitions to quantum gravity.

During the early development of CFT, a pinnacle was the discovery that the operator product expansion (OPE) in a class of two-dimensional theories leads to certain null states, resulting in a large number of differential equations that correlators with degenerate fields obey \cite{BELAVIN1984333}. These equations allow the correlators to be computed exactly, without using Lagrangian perturbation methods. In general dimensions, holographic duality \cite{Maldacena:1997re, Gubser:1998bc, Witten:1998qj} and the conformal bootstrap (see, $e.g.$, \cite{Poland:2018epd} for a review) are powerful frameworks for analyzing strongly-coupled conformal observables. However, the null-state-type equation approach is absent in higher dimensions. Our working hypothesis is that, despite the increased complexity of the OPE in higher dimensions, similar governing equations may emerge in a class of CFTs with gravity duals.

In recent years, there has been very fruitful progress in analyzing the analytic structure of conformal correlators in higher dimensions using holography and bootstrap-related methods. A prototypical example is a scalar two-point correlator at finite temperature, which can be computed by solving a scalar field equation in an Anti-de Sitter (AdS) black hole background. The thermal correlator may be viewed as the scattering amplitude of a light probe scalar in a heavy background, making it equivalent to a heavy-light four-point scalar correlator at large central charge (large N).  In holographic theories, there are two types of contributions. The first arises from the stress-tensor sector, involving multi-stress tensor operators that are primary stress-tensor composites. The second involves double-trace operators constructed from the scalars. Our focus here will be on the multi-stress tensor contributions. In two-dimensional CFTs, the full resummation of multi-stress tensors is encapsulated by the Virasoro identity block, which can be computed via the Belavin-Polyakov-Zamolodchikov (BPZ) differential equations; see, $e.g.$, \cite{zamolodchikov1984conformal, zamolodchikov2, Harlow:2011ny, Hartman:2013mia, Fitzpatrick:2014vua, Perlmutter:2015iya, Chen:2016cms, Besken:2019jyw}. The main goal of this work is to derive BPZ-type equations in higher-dimensional CFTs via holography. We will concentrate on four dimensions as the higher-dimensional example.

Some multi-stress tensor OPE coefficients in four dimensions were extracted from a dual gravity computation \cite{Fitzpatrick:2019zqz}. A lesson, perhaps surprising, is that the stress-tensor sector of the theory can be determined through a near-boundary analysis, while the double-trace operators depend on a boundary condition near the black hole horizon. For related holographic computations, see, $e.g.$, \cite{Kulaxizi:2018dxo, Li:2019tpf, Kulaxizi:2019tkd, Fitzpatrick:2019efk, Parnachev:2020fna, Fitzpatrick:2020yjb, Alday:2020eua, Grinberg:2020fdj, Rodriguez-Gomez:2021mkk, Krishna:2021fus,  Karlsson:2022osn, Dodelson:2022yvn, Huang:2022vet, Dodelson:2023vrw, Esper:2023jeq, Buric:2025anb, Buric:2025fye, Bajc:2025jjv, Barrat:2025twb, Giombi:2026kdz, Arnaudo:2026der, Jia:2026ryl}. Recent progress has further demonstrated that the stress-tensor sector of boundary conformal theories can probe black hole singularities in higher dimensions; see \cite{Ceplak:2024bja, Afkhami-Jeddi:2025wra, Ceplak:2025dds, Araya:2026shz, Grozdanov:2026cut} and \cite{Giombi:2026kdz, Arnaudo:2026der, Jia:2026ryl}. Some earlier works include \cite{Louko:2000tp, Kraus:2002iv, Fidkowski:2003nf, Brecher:2004gn, Hamilton:2005ju, Festuccia:2005pi}. 
The quest to understand the interior structure of black holes thus provides additional motivation to seek greater analytic control over multi-stress tensor contributions in holographic theories.      
On the other hand, field-theoretic methods have been employed to derive several expressions that sum over spins, for double- and triple-stress tensors, based on the bootstrap approach \cite{Karlsson:2019dbd, Karlsson:2020ghx}. Related CFT techniques applicable to higher dimensions were also developed in, $e.g.$, \cite{Iliesiu:2018fao, Li:2019zba, Karlsson:2021duj, Parisini:2022wkb, Parisini:2023nbd, Haehl:2025ehf, Niarchos:2025cdg}. Given these recent developments and the available multi-stress tensor OPE coefficients in higher-dimensional CFTs, it seems natural to look for potential organizing differential equations. 

Inspired by the analogy with $d=2$ BPZ equations at large central charge, a set of linear differential equations was proposed in \cite{Huang:2023ikg} to organize the minimal-twist multi-stress tensor contributions, to all orders in the number of exchanged stress tensors; see \cite{Huang:2024wbq} for analysis of their exact solutions. These conjectured equations have passed several non-trivial checks: their solutions are consistent with all known available OPE coefficients, including the resummed structures for double- and triple-stress tensors in four dimensions obtained in \cite{Kulaxizi:2019tkd, Karlsson:2019dbd, Karlsson:2020ghx}. While the field-theoretic mechanism underlying these equations remains elusive, in this work we derive, and thereby confirm, their existence through holography. We hope this progress will aid in developing field-theoretic methods for computing correlators in higher dimensions, as well as lead to further insights into black hole singularities and the holographic framework more generally.

We shall begin with a warm-up computation in the AdS$_3$/CFT$_2$ case in the next section. Starting from a bulk scalar equation in the Banados-Teitelboim-Zanelli (BTZ) black hole geometry \cite{Banados:1992wn}, we will derive the (level-2) BPZ equation in the dual two-dimensional CFTs at large central charge, where the degenerate light scalars in the correlator have conformal dimension $\Delta= 2h = -1 + {\cal  O}({1\over c})$.  To our knowledge, even in this low-dimensional case, establishing a connection between the classical bulk equation and the boundary null-state equation is a new result. The main focus of the present work will be the AdS$_5$/CFT$_4$ case, presented in Section 3, where we will show similar BPZ-type equations emerge from gravity.  We will adopt a black hole with a spherical horizon, enabling us to capture a more general structure than that of a planar black hole. By performing a near-boundary expansion of the bulk field equation, we observe that, order by order, the higher-order perturbative solutions decouple when the conformal dimensions of the probe light scalar take certain (negative) integer values. The decoupling reduces the problem to a set of lower-order equations. These equations, valid  for fixed conformal dimensions, are interpreted as differential equations in the dual CFTs. In particular, the results agree with the three $d=4$ CFT equations conjectured in \cite{Huang:2023ikg} that correspond to the $\Delta = -1, -2, -3$ cases focusing on the minimal-twist multi-stress tensors. The holographic calculation can yield more CFT equations, including those involving other degenerate-like scalars, $e.g.$, for $\Delta = -4$, and equations not restricted to minimal-twist operators. These results hint at the existence of a vast number of analogous equations in higher-dimensional CFTs with gravity duals. In Section 4, we conclude with some open questions for future investigation.

\vspace{0.2cm}
\noindent {\bf BPZ from BTZ:}  
As the simplest case, we will first derive the celebrated (level-2) BPZ null-state equation starting from a bulk scalar equation in 
AdS$_3$ with a spherical black hole. We adopt the following Euclidean metric: 
\begin{align}
ds^2= \big(1+ r^2 f(r) \big) dt^2+ {dr^2 \over 1+ r^2 f(r) } + r^2 d\theta^2
\end{align}  
with $f(r) = 1- {\mu \over r^2}$ where the AdS radius is set to 1. The parameter $\mu$ is proportional to the black hole mass, and, according to AdS/CFT, it relates to the ratio between the weight of a heavy operator and the central charge.  A common notation uses
$\mu = 24 \eta$ where  $\eta  \equiv  {h_H \over c}$ is fixed, with $h_H$ and the central charge $c$ taken to be infinitely large. 
To study the conformal correlator, we analyze the bulk scalar equation, $\left( - \nabla^2 + m^2 \right) \Phi (t, r, \theta) = 0$, which can be written as  
\begin{align}
\label{3DbulkEoM}
m^2 \Phi & = {1\over 1 + r^2 - \mu}\,\Phi_{tt}\\
&
+ {1\over r^2} \Big( \Phi_{\theta\theta}
+ r (1 + 3 r^2 - \mu)\,\Phi_r
+ r^2 (1 + r^2 - \mu)\,\Phi_{rr}\Big) \nn \ .
\end{align} 
The mass of the bulk scalar is related to the dimension of the boundary scalar via $m^2= \Delta (\Delta - 2) $. We use $\Delta=2h$.   
The conformal correlator can be obtained by the boundary limit, $\lim_{r \to \infty} r^{\Delta} \Phi (t, r, \theta)$.  
The pure AdS solution is \cite{Witten:1998qj} 
\begin{align}
\label{puresol}
\Phi (t, r, \theta)_{\mu=0}= \Big({1\over 2\left(\sqrt{1 + r^2}\,\cosh t - r\cos\theta\right)}\Big)^{\Delta} \ .
\end{align} 
In the BTZ background, the closed-form bulk-to-boundary propagator was obtained in \cite{Keski-Vakkuri:1998gmz}. However, since closed-form expressions in higher dimensions become intractable, we shall develop an approach that can be generalized to higher dimensions.

Here we consider a near-boundary expansion:
 \begin{align}
\label{nearbry}
 \Phi= r^{-\Delta}\left( \Phi_{0}(t,\theta)
+ {\Phi_{2}(t,\theta)\over r^{2}}
+ {\Phi_{4}(t,\theta)\over r^{4}}
 + {\cal O} (r^{-6}) 
\right) \ .
\end{align} 
Plugging this into \eqref{3DbulkEoM}, the leading order yields the relation 
$m^2 = (\Delta - 2)\Delta$. The subleading order equation reads 
 \begin{align}
 4 \Delta\,\Phi_{2} = \Delta^2 (\mu - 1)\,\Phi_{0} - \Phi_{0,tt}
- \Phi_{0,\theta\theta} \ .
\end{align}  
Since the dynamical structure trivializes when $\Delta=0$, we will not consider that case. 
After solving for $\Phi_{2}$ in terms of $\Phi_{0}$, the next-order equation is given by
 \begin{align}
\label{3Ddecouple}
32\,\Delta(\Delta+1)\,\Phi_{4} &= \Big[
(\partial_{t}^{2}+\partial_{\theta}^{2})^{2} + \alpha \partial_{t}^{2}+\beta \partial_{\theta}^{2}+\gamma \Big]\Phi_{0}
\end{align} 
where
 \begin{align}
\alpha &= - 2 \Delta(\Delta+4)(\mu-1) - 4(\mu-1) \ , \nn\\
 \beta &= - 2 \Delta(\Delta+2)(\mu-1)  -4 (\mu-1) \ ,  \\ 
\gamma &= \Delta^{2}(\Delta+2)^{2}(\mu-1)^{2} \ . \nn
\end{align} 
Here we see that $\Phi_{4}$ decouples at $\Delta= -1$ (or equivalently, $h= - {1\over 2}$), which is precisely the known value for the level-2 BPZ degenerate field at large central charge. The resulting equation \eqref{3Ddecouple} at $\Delta= -1$ can be viewed as a boundary CFT differential equation, where $\Phi_{0}$ corresponds to a thermal two-point correlator, or a heavy-light four-point correlator. 
Let us perform a coordinate transformation to the $z, \zb$ coordinates via 
 \begin{align}
\label{map}
(1-z)= e^{t + i \theta} \ , ~~~ (1-\zb)= e^{t - i \theta}  \ .
\end{align} 
Note the conformal mapping generates an overall factor in the correlator: 
$\Phi_{0} (t,\theta) \to  {\big((1-z) (1- \zb)\big)}^{\Delta\over 2} \Phi_{0} (z,\zb)$.   
Considering that the correlator factorizes into holomorphic and anti-holomorphic parts, $\Phi_{0} (z,\zb) = \Phi_{0} (z)  {\bar \Phi}_{0} (\zb)$, 
the equation \eqref{3Ddecouple} at $\Delta= -1$ reduces to the following decoupled equations:
 \begin{align}
&  (1 - z)^2\,\partial^2  \Phi_{0}(z) + 6\eta\,\Phi_{0}(z) = 0 \ , \\
&   (1 - \bar z)^2\, \bar\partial^2 \bar \Phi_0(\bar z) + 6\eta\, \bar \Phi_{0}(\bar z) = 0 \ .
\end{align} 
where we recalled $\mu=24 \eta$. These are the BPZ null-state equations in $d=2$ CFTs at large central charge.

\vspace{0.2cm}
\noindent {\bf A $d=4$ CFT equation from Einstein gravity:}  
The approach used in the BTZ case can be extended to higher dimensions. 
We focus on $d=4$ holographic CFTs and consider an AdS$_5$-Schwarzschild spherical black hole: 
 \begin{align}
ds^2 = (1 + r^2 f(r))\,dt^2
+ \frac{dr^2}{1 + r^2 f(r)}
+ r^2  d\Omega^2_3
\end{align} 
where $f(r) = 1- {\mu \over r^4}$ with the AdS radius set to unity. The unit 3-sphere, $d\Omega^2_3$, uses angular coordinates $\theta_1, \theta_2, \theta_3$. Due to the rotational symmetry, we will drop the dependence on $\theta_2, \theta_3$ and denote $\theta_1=\theta$ in the bulk scalar equation, which can be written as 
 \begin{align}
 m^2 \Phi & = \frac{r^2}{r^2 + r^4 - \mu}\,\Phi_{tt} + \frac{1}{r^3} \Big(  r\,\Phi_{\theta \theta}+ 2r\cot\theta\,\Phi_{\theta} \nn\\
&+ (3r^2 + 5r^4 - \mu)\,\Phi_r + r(r^2 + r^4 - \mu)\,\Phi_{rr} \Big) \ .
\end{align} 
The pure AdS solution takes the same form as \eqref{puresol}. 

We adopt a near-boundary expansion:
 \begin{align}
 \Phi = r^{-\Delta} \sum_{i =0, 2, 4, 6, \dots}  {\Phi_{i}(t,\theta)\over r^i} \ . 
\end{align} 
The leading order equation yields the standard relation $m^2= \Delta (\Delta - 4)$ where $\Delta$ is the conformal dimension of the dual CFT scalar. The subleading order equation is 
 \begin{align}
 4(1-\Delta ) \Phi_{2}=
\Delta (\Delta - 2) \Phi_{0}
+2\cot\theta \, \Phi_{0,\theta}
+\Phi_{0,\theta\theta}
+\Phi_{0,tt} \ .
\end{align} 
When $\Delta = 1$, the mass becomes imaginary, and additionally, this equation has no $\mu$ dependence. We will move to the next order. Solving for $\Phi_{2}$ in terms of $\Phi_{0}$, 
we substitute back into the next-order equation, which now involves only $\Phi_4$ and $\Phi_0$.   This equation is straightforward to derive, and for brevity we will not present it here.  
The key difference, compared with the AdS$_3$ case, is that there is no way to decouple $\Phi_4$ for a non-zero $\Delta$. So, we proceed further to the next order. 

After expressing $\Phi_4$ in terms of $\Phi_{0}$, the subsequent order equation involves only $\Phi_6$ and $\Phi_0$. We find that the coefficient of  $\Phi_6$ is proportional to
 \begin{align}
\label{phi6condi}
\Phi_6: (\Delta+1) \ .
\end{align} 
Thus, $\Phi_6$ decouples at $\Delta= - 1$.    
We perform the transformation \eqref{map} to the $z, \zb$ coordinates.   
To express the equation concisely, it is convenient to let 
 \begin{align}
\label{m1QF}
\Phi_{0, \Delta = -1}(z,\zb) = \frac{1-\bar z}{z\,\bar z\,(z-\bar z) (z\,\bar z)^{\Delta}}\, Q_{\Delta = -1}  (z,\zb)  
\end{align} with $\Delta= -1$. We adopt $(z\,\bar z)^{\Delta}$ for later convenience.
The full CFT equation is  
 \begin{align}
\label{4dBPZfull}
&0=\Big[ \, 
(1-z)^3(1-\bar z)^3 \partial^3\bar\partial^3
- 3 (1-z)^3(1-\bar z)^2 \,\partial^3\bar\partial^2  \nn\\
&~~~~~~~~ + \mu (1-z)^2 \partial^2+\mu (1-z) (1-\bar z) \partial\bar\partial 
  \nn\\
&~~~~~~~~ +  \mu (1-\bar z)^2 \bar\partial^2 + \mu (1-z) \partial  \, \Big] \, Q_{\Delta = -1} (z,\zb)  \ .
\end{align} 

The equation \eqref{4dBPZfull} can be simplified by focusing on the lightcone limit, where minimal-twist operators dominate. The near-lightcone dynamics is of particular interest in the context of analytic conformal bootstrap \cite{Fitzpatrick:2012yx, Komargodski:2012ek}, and this also allows us to compare to the $d=4$ differential equations recently conjectured in \cite{Huang:2023ikg}.  
To isolate the contribution from minimal-twist operators, recall the near-lightcone behavior of the $d=4$ conformal block \cite{Dolan:2000ut}: 
\begin{align}
 (z \bar{z})^{\frac{\tau}{2}} \left( -\frac{z}{2} \right)^J {}_2F_1(\frac{\tau}{2}+J, \frac{\tau}{2}+J, \tau+2J, z) + {\cal O}(\bar{z}^{\frac{\tau}{2}+1}) \ .
\end{align} 
Here $\tau$ and $J$ are twist and spin, respectively.  When $\zb \to 0$, which is formal in the Euclidean but corresponds to the lightcone limit in Lorentzian signature, higher-twist operators are suppressed.  
To prevent a vanishing result due to the factor $\bar{z}^{\tau\over 2}$, we set 
\begin{align}
v= \mu \zb  ~~  {\rm  fixed}
\end{align} 
while taking $\zb \to 0$. 
After isolating the minimal-twist contribution, the explicit $\mu$ dependence can be restored in the equation.  
In this limit, we find \eqref{4dBPZfull} reduces to
 \begin{align}
\label{eqQ1}
\bar\partial^2 \big(\mu+ (1 - z)^3\,\partial^3 \bar\partial\big)\,Q^{\tau_{\rm min}}_{\Delta = -1} = 0 \ .
\end{align} 
In the same limit, the relation \eqref{m1QF} simplifies to
 \begin{align}
(z\,\bar z)^{\Delta} \, \Phi^{\tau_{\rm min}}_{0, \Delta = -1} = \frac{1}{z^2\,\bar z} \,  Q^{\tau_{\rm min}}_{\Delta = -1} \equiv F^{\tau_{\rm min}}_{\Delta = -1} 
\end{align} 
where $F=F(z, \zb)$ is introduced for later convenience.   
The equation \eqref{eqQ1} can be written as 
 \begin{align}
\label{Feq0}
 \Big[
 \left(1 + \bar z\,\bar\partial\right)
\big(
 \partial^3
+ \frac{6}{z}\partial^2
+\frac{6}{z^2}\partial
\big)&+ \frac{\mu \zb}{(1-z)^3}
\Big]\,F^{\tau_{\rm min}}_{\Delta = -1} \nn\\
&= \alpha(z)+\beta(z) \zb  
\end{align}  
where $\alpha(z)$ and $\beta(z)$ come from removing the derivatives $\bar{\partial}^2$ acting globally in \eqref{eqQ1}. 
To determine these functions, we require that $F(z,\zb)$ admits the following expansion (without restricting to $\Delta= -1$):
 \begin{align}
\label{lightconeF}
F^{\tau_{\rm min}}(z, \zb) = \sum_{k=0, 1, 2, 3, \dots}^{\infty} g_k(z) \, (\mu \zb)^k \ .
\end{align} 
In particular, $g_0=1$ corresponds to the identity contribution, and $g_1= \frac{\Delta}{120}\, f(3, z)$ with $f(a,z) = z^{a}\,{}_2F_1(a,a;2a;z)$ is 
the minimal-twist contribution from the global block, associated with the single-stress tensor exchange universally fixed by the Ward identities \cite{Dolan:2000ut}.  
Contributions for $k>1$ represent $k$-stress tensors with a summation over spins. 
 We find the consistency with $g_0$ and $g_1$ fixes $\alpha(z) = \beta(z) = 0$. 
The resulting equation, for all minimal-twist multi-stress tensors, reads
 \begin{align}
\Big[
 \left(1 + \bar z\,\bar\partial\right)
\big(
 \partial^3
+ \frac{6}{z}\partial^2
+\frac{6}{z^2}\partial
\big)+ \frac{\mu \zb}{(1-z)^3}
\Big]\,F^{\tau_{\rm min}}_{\Delta = -1}   =0  \ .
\end{align}  This equation, equivalent to $\big(\mu+ (1 - z)^3\,\partial^3 \bar\partial\big)\,Q^{\tau_{\rm min}}_{\Delta = -1} =0$, first appeared in \cite{Huang:2023ikg}. 
It was proposed based on guesswork and has passed some consistency checks. Here, we have derived it directly from Einstein gravity.

\vspace{0.2cm}
\noindent {\bf More $d=4$ CFT equations:}  The derivations for the $d=4$ CFT equations at other special values of $\Delta$ follow a similar pattern. If we do not set $\Delta = -1$ at the $\Phi_6$ level, $i.e.$, \eqref{phi6condi}, but instead solve for  $\Phi_6$ in terms of $\Phi_0$, and then substitute it into the next-order equation, we find that the coefficient of $\Phi_8$ is proportional to
 \begin{align}
\label{phi8condi}
\Phi_8: (\Delta+2) \ . 
\end{align} 
One can proceed in the same way as the previous case to obtain a CFT equation at $\Delta = - 2$.   
The full equation is lengthy and will not be listed explicitly here. A simplification occurs in the minimal-twist limit, which yields 
 \begin{align}
\label{4dQm2}
\bar\partial^3\Big[
9\mu
+ 6\mu(1-z)\,\partial
+ (1-z)^4\,\partial^4 \bar\partial
\Big]\,Q^{\tau_{\rm min}}_{\Delta = -2} = 0 
\end{align}  
with $(z\,\bar z)^{\Delta} \, \Phi^{\tau_{\rm min}}_{0, \Delta = -2} = \frac{1}{z^3\,\bar z^2} \,  Q^{\tau_{\rm min}}_{\Delta = -2}$. 
Removing the overall derivatives $\bar\partial^3$ leads to three undetermined functions.  
As in the previous case, the consistency with the single-stress tensor block (and the identity operator) forces these functions to vanish. The resulting equation, $\big(9\mu+ 6\mu(1-z)\,\partial+ (1-z)^4\,\partial^4 \bar\partial\big) Q^{\tau_{\rm min}}_{\Delta = -2} = 0$, matches the  previously conjectured equation in \cite{Huang:2023ikg}. 

Let us extend the computation to the next two orders. Using the solutions from previous orders, we find that the coefficients of $\Phi_{10}$ and $\Phi_{12}$ are proportional to
 \begin{align}
 \Phi_{10}: (\Delta+3) \ , ~~~~~ \Phi_{12}: (\Delta+4) \ .
\end{align}  Applying the same decoupling mechanism, the equations at $\Delta= -3, -4$ can be obtained. (One can also see that the pattern is $\Phi_{2n}: (\Delta + n - 2)$ with $n = 3, 4, 5, 6, \dots$) 
It is illuminating to present the corresponding equations for minimal-twist operators. We obtain  
 \begin{align}
\label{4dQm3}
\bar\partial^4\Big[
72\mu
&+ 63 \mu (1-z) \partial    \\
&+21 \mu (1-z)^2 \partial^2 + (1-z)^5\,\partial^5 \bar\partial
\Big]\,Q^{\tau_{\rm min}}_{\Delta = -3}=0  \nn
\end{align} 
 \begin{align}
\label{4dQm4}
&\bar\partial^4\Big[
100\mu^2 + 600\mu\,\bar\partial
+ 576\mu(1-z)\partial\bar\partial + 252\mu(1-z)^2\partial^2\bar\partial \nn\\
& 
+ 56\mu(1-z)^3\partial^3\bar\partial+ (1-z)^6\partial^6\bar\partial^2
\Big] \,Q^{\tau_{\rm min}}_{\Delta = -4}=0 
\end{align} 
where we introduce $(z\,\bar z)^{\Delta} \, \Phi^{\tau_{\rm min}}_{0,\Delta = -3} = \frac{1}{z^4\,\bar z^3} \,  Q^{\tau_{\rm min}}_{\Delta = -3}$ and  $(z\,\bar z)^{\Delta} \, \Phi^{\tau_{\rm min}}_{0,\Delta = -4}= \frac{1}{z^5\,\bar z^4} \,  Q^{\tau_{\rm min}}_{\Delta = -4}$ to simplify the expressions. The universal single-stress tensor block again allows us to remove the overall $\bar\partial^4$ from both equations with a vanishing integration constant.  The resulting CFT equation obtained from \eqref{4dQm3}, $i.e.$, $\big(72\mu+ 63 \mu (1-z) + 21 \mu (1-z)^2 \partial^2 \partial + (1-z)^5\,\partial^5 \bar\partial \big) Q^{\tau_{\rm min}}_{\Delta = -3}=0$, matches the proposed minimal-twist equation in \cite{Huang:2023ikg}. On the other hand, equation \eqref{4dQm4} is new. As a consistency check, we have verified that, using the expansion \eqref{lightconeF} with the boundary conditions $g_i(0)=0$ for $i\neq 0$ (see \cite{Huang:2023ikg, Huang:2024wbq} for more discussions on solving this kind of equations), it reproduces the resummed contributions from double- and triple-stress tensors computed in \cite{Kulaxizi:2019tkd, Karlsson:2019dbd, Karlsson:2020ghx}.

Let us also remark that the equations proposed in \cite{Huang:2023ikg} for $\Delta = -1, -2, -3$ are based on pattern recognition. This raises a question: does a pattern discontinuity occur at $\Delta = -4$? Comparing \eqref{4dQm4} with the previous equations, $i.e.$, \eqref{eqQ1}, \eqref{4dQm2}, \eqref{4dQm3}, the pattern related to $\bar{\partial}$ is disrupted. Understanding this pattern discontinuity from the field-theory perspective could be interesting, but the holographic approach provides a systematic way to generate the CFT equations at these special $\Delta$s. It would be valuable to develop an algorithm to systematically produce the $d=4$ CFT differential equations for any negative integer $\Delta$. By performing an analytic continuation of the  correlators involving these degenerate-like fields, one might gain insight into the structure of the $d=4$ correlators for general conformal dimensions.

\vspace{0.2cm}
\noindent {\bf Conclusion:} We have provided a holographic derivation of BPZ-type differential equations in higher-dimensional CFTs at large central charge. The gravity calculation confirms the previously conjectured equations for minimal-twist multi-stress tensors and reveals additional equations. These results suggest that a vast number of differential equations may exist in higher-dimensional CFTs with gravity duals. We expect the computation can be extended to other spacetime dimensions.

It would be worthwhile to investigate whether the analytic solutions to these equations could provide insights into the interior of black holes. Our approach adopts a near-boundary expansion and probing regions farther from the boundary would require a large negative conformal dimension. Relatedly, it is natural to consider higher-curvature corrections, either through a minimally or non-minimally coupled bulk scalar, to see if the special values of the conformal dimension may be altered, and to obtain stringy corrections to the differential equations.

Having derived the CFT equations from gravity, it would be nice to understand these equations via a field-theoretic mechanism. Could these higher-dimensional equations be rooted in certain large-N CFT symmetries?

\vspace{0.2cm}

I would like to thank T. Hartman and M. Rangamani for the suggestion to look for a potential connection between the $d=4$ CFT equations and the bulk equation at an early stage of this work. I also thank M. Banados for helpful correspondence and F. Haehl for related discussions. This work was supported in part by UKRI under the Horizon Europe Funding Guarantee EP/X030334/1.      

\vspace{-0.5cm}
\bibliographystyle{utphys}  
\bibliography{BPZfromG}  


\end{document}